\documentclass[a4paper,fleqn,usenatbib,useAMS]{mnras}

\usepackage{color}
\usepackage{graphicx}
\usepackage{amsmath}	
\usepackage{amssymb}	
\usepackage{multicol}        
\usepackage{bm}		
\usepackage{pdflscape}	
\usepackage{subcaption}
\usepackage{comment}

\usepackage[T1]{fontenc}
\usepackage{ae,aecompl}
\usepackage{enumitem}
\usepackage{colortbl}

\usepackage{newtxtext,newtxmath}

\usepackage{etoolbox}

\title{The origin of the black hole offset in M31} 

\author[P. Boldrini]{Pierre Boldrini$^{1}$\thanks{Contact e-mail: \href{mailto:boldrini@iap.fr}{boldrini@iap.fr}}
\\
$^{1}$Sorbonne Universit\'e, CNRS, UMR 7095, Institut d'Astrophysique de Paris, 98 bis bd Arago, 75014 Paris, France}
\date{In original form 2020 February 29.}

\pubyear{2020}

\begin{document}
\label{firstpage}
\pagerange{\pageref{firstpage}--\pageref{lastpage}}
\maketitle

\begin{abstract}

Using state-of-the-art high-resolution fully GPU N-body simulations, we demonstrate for the first time that the infall of a dark matter rich satellite naturally explains a present black hole offset by sub-parsecs in M31. Observational data of the tidal features provide stringent constraints on the initial conditions of our simulations. The heating of the central region of M31 by the satellite via dynamical friction entails a significant black hole offset after the first pericentric passage. After having reached its maximum offset, the massive black hole sinks towards the M31 centre due to dynamical friction and it is determined to be offset by sub-parsecs as derived by observations.

\end{abstract}





\section{Introduction}

Most galaxies are known to harbour supermassive black holes (SMBHs), formed within a billion years after the Big Bang. They reside in the centres of present day galaxies with masses of $\sim 10^{6}-10^{10}$ M$_{\sun}$ based on observations of high-redshift quasars \citep[see][for a review]{2013ARA&A..51..511K}. Intriguingly, some observations of active galactic nuclei claim that massive black holes (MBHs) are not located at the centres of their host galaxies \citep{2014ApJ...796L..13M,2016ApJ...817..150M,2019arXiv190904670R,2019ApJ...885L...4S}. Different scenarios have been proposed to explain these off-centered BHs. Plausibly, the offset could be due to the presence of a binary system before the merger \citep[e.g.][and references therein]{2010PhRvD..81j4009S}, or via tidal stripping during mergers \citep[see][and references therein]{2018ApJ...857L..22T}, the incomplete MBH inspiralling phase of the two merging galaxies \citep{2009ApJ...690.1031B,2014ApJ...789..112C}, or the recoil of merging BHs \citep{2005LRR.....8....8M,2005MNRAS.358..913V,2007PhRvL..99d1103L,2012AdAst2012E..14K}. It was pointed out that the majority of off-centered BHs are present in host galaxies showing signs of interactions/mergers \citep{2019arXiv190904670R}.

One of the most striking features of the Andromeda galaxy is the presence of a doubled-peak nucleus in the central region. These two peaks, P1 and P2, are separated by 0".49 corresponding to a projected distance of 1.8 pc at the distance of M31 \citep{1993AJ....106.1436L,1999ApJ...522..772K}. The optically faint peak P2 has been identified as hosting 
a MBH of mass $1.5\times10^{8}$ M$_{\sun}$.  \citep{1995ARA&A..33..581K,1999ApJ...522..772K,2009ApJ...698..198G}. It was shown that the M31 BH is offset by 0.26 pc from P2 considered as the centre of the galaxy \citep{1999ApJ...522..772K}. As there are several strong indications of a recent merger activity in M31, we propose below a new explanation for this off-centre MBH.

M31 is predicted to arise from the merger and accretion of many smaller sub-systems \citep{1978MNRAS.183..341W,1991ApJ...379...52W}. This hypothesis is supported by the discovery of tidal features such as giant stellar stream (GSS) to its south as well as giant stellar shells to the east and west of its centre \citep{2001Natur.412...49I,2004MNRAS.351..117I,2005ApJ...634..287I,2002AJ....124.1452F,2003MNRAS.343.1335M,2006AJ....131.2497G,2008ApJ...689..958K}. It is widely believed that the phased features of M31 result from the accretion of a satellite galaxy \citep{2004MNRAS.351..117I,2006AJ....131.1436F,2006MNRAS.366.1012F,2007MNRAS.380...15F}. After examining the orbits and the mass of an accreting satellite galaxy, numerous high-resolution $N$-body simulations have been extremely successful in reproducing these structures \citep{2006AJ....131.1436F,2006MNRAS.366.1012F,2007MNRAS.380...15F,2013MNRAS.434.2779F,2008ApJ...674L..77M,2010ApJ...725..542H,2014PASJ...66L..10K,2014ApJ...783...87M,2014MNRAS.442..160S,2016ApJ...827...82M,2017MNRAS.464.3509K}.

In this Letter, we show that the accreting satellite, as the origin of the GSS and stellar shells, heated the central region of M31 and kicked the central MBH from the galaxy centre. Assuming the cosmologically plausible scenario from \cite{2014MNRAS.442..160S} for the satellite, we performed state-of-the-art $N$-body simulations with GPUs, which allow parsec resolution, to study this heating process that naturally explains the present BH offset in M31. The letter is organized as follows. Section 2 provides a description of the $N$-body modelling of M31 and its satellite, along with details of our numerical simulations. In Section 3, we present our simulation results and discuss the origin of the BH offset in M31. Section 4 presents our conclusions.

\section{High-resolution fully GPU N-body simulation}

The initial conditions for the M31 satellite are taken from \cite{2014MNRAS.442..160S} (see details in Table~\ref{tab1}). The dark matter rich satellite starts at its first turnaround radius at ($x_0$,$y_0$,$z_0$)=(-84.41,152.47,-97.08) with a null velocity in a reference frame centered on M31 with the x-axis pointing east, the y-axis pointing north and the z-axis corresponding to the line-of-sight direction. We add a massive BH with a mass of $1.5\times10^{8}$ M$_{\sun}$ as a point mass in the center of M31 \citep{2009ApJ...698..198G}. To generate our live objects, we use the initial condition code \textsc{magi} \citep{2017arXiv171208760M}. Adopting a distribution-function-based method, it ensures that the final realization of the galaxy is in dynamical equilibrium \citep{2017arXiv171208760M}. We perform our simulations with the high performance collisionless $N$-body code \textsc{gothic} \citep{2017NewA...52...65M}. This gravitational octree code runs entirely on GPU and is accelerated by the use of hierarchical time steps in which a group of particles has the same time step \citep{2017NewA...52...65M}. We evolve the M31 galaxy-satellite system over 2.5 Gyr in each scenario. We set the particle resolution of all the live objects to $4.4\times10^{4}$ M$_{\sun}$ and the gravitational softening length to 2 pc.

\begin{table}
\begin{center}
\label{tab:landscape}
\begin{tabular}{cccccccccccc}
 \hline
   Component & Profile & a & $r_{200}$ &  Mass \\
     & & [kpc] & [kpc] & [$10^{10}M_{\sun}$] \\
    \hline
   M31 halo & NFW & 7.63 & 195 & 88 \\
   M31 bulge & Hernquist & 0.61 & - & 3.24 \\
   M31 disk & Exponential & $R_{\mathrm{d}}=$ 5.4 & - & 3.66\\
   & disk & $z_{\mathrm{d}}=$ 0.6 & - & - \\
   M31 black hole & Point mass & - & - & 0.015 \\

    \hline
   Satellite halo & Hernquist & 12.5 & 20 & 4.18 \\
   Satellite stars & Plummer & 1.03 & - & 0.22  \\
    \hline
\end{tabular}
\caption{{\it Simulation parameters:} From left to right, the columns provide for each component: the density profile, the scale length, the virial radius, the mass. We set the initial positions in a reference frame centered on M31 with the x-axis pointing east, the y-axis pointing north and the z-axis corresponding to the line-of-sight direction. We consider an infalling scenario of a dark matter rich satellite \citep{2014MNRAS.442..160S} where the satellite starts at its first turnaround radius at ($x_0$,$y_0$,$z_0$)=(-84.41,152.47,-97.08) with a null velocity. We set the particle resolution of all the live objects to $4.4\times10^{4}$ M$_{\sun}$ and the gravitational softening length to 2 pc. We also add a massive BH as a point mass in the center of M31 with a mass of $1.5\times10^{8}$ M$_{\sun}$ \citep{2009ApJ...698..198G}.}
\label{tab1}
\end{center}
\end{table}

\section{Results}

\begin{figure}
\centering
\includegraphics[width=0.47\textwidth]{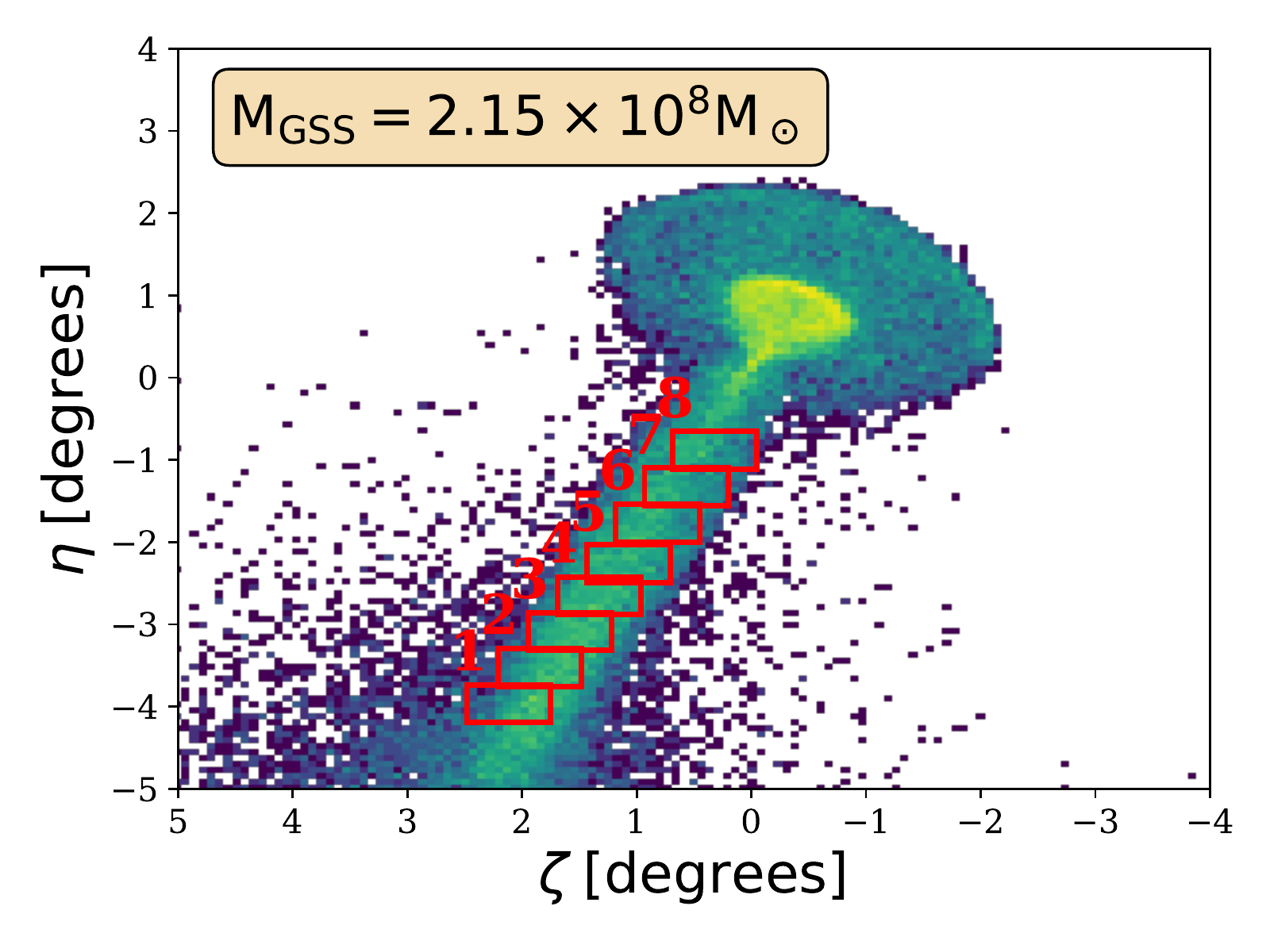}
\caption{{\it Giant south stream of M31:} Simulated stellar density maps in standard sky coordinates corresponding to particles of satellite stars at 2.1 Gyr. We represent the observed stream fields by black boxes \protect{\citep{2003MNRAS.343.1335M}}. We find $M_{\mathrm{GSS}}=2.15\times10^8M_{\sun}$ in good agreement with the value of $2.4\times10^8M_{\sun}$ derived from observations with a mass-to-light ratio of 7 \citep{2001Natur.412...49I,2006MNRAS.366.1012F}}
\label{fgR0}
\end{figure}

\begin{figure}
\centering
\includegraphics[width=0.47\textwidth]{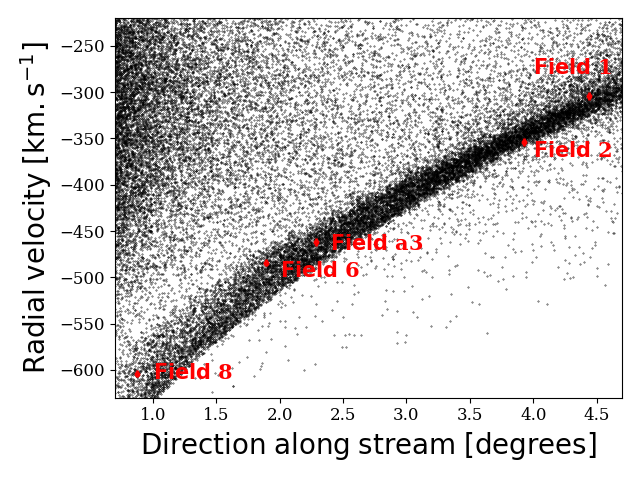}
\caption{{\it Comparison with kinematic data of the observed GSS:} Simulated radial velocity of satellite particles as a function of the distance along the stream at 2.1 Gyr. We represent the radial velocity measurements in five fields by red points with error-bars \citep{2004MNRAS.351..117I,2006MNRAS.366.1012F}. A good agreement with observations for the radial velocity measurement is shown.}
\label{fgR1}
\end{figure}

\subsection{Comparison with M31 observations}

First, we assess our model by making a detailed comparison with M31 observations in this scenario invoking the infall of a dark matter rich satellite (see Table.~\ref{tab1}). Fig.~\ref{fgR0} depicts the simulated stellar density maps in standard sky coordinates corresponding to particles of satellite stars at 2.1 Gyr. We represent the observed stream fields as solid rectangles with proper scaling. We note that the simulated stream is in good agreement with the observations regarding the morphology and spatial extent of the GSS. We find $M_{\mathrm{GSS}}=2.15\times10^8$ M$_{\sun}$ in good agreement with the value of $2.4\times10^8 $ M$_{\sun}$ derived from observations with a mass-to-light ratio of 7 \citep{2001Natur.412...49I,2006MNRAS.366.1012F}. Furthermore, we test the infalling model of \cite{2014MNRAS.442..160S} against kinematic data. Fig.~\ref{fgR1} shows simulated radial velocities of satellite particles as a function of the distance along the stream. We obtain a good agreement with observations for the radial velocity measurement in the five fields \citep{2004MNRAS.351..117I,2006MNRAS.366.1012F}.

\subsection{Off-centre MBH in M31}

\begin{figure}
\centering
\includegraphics[width=0.47\textwidth]{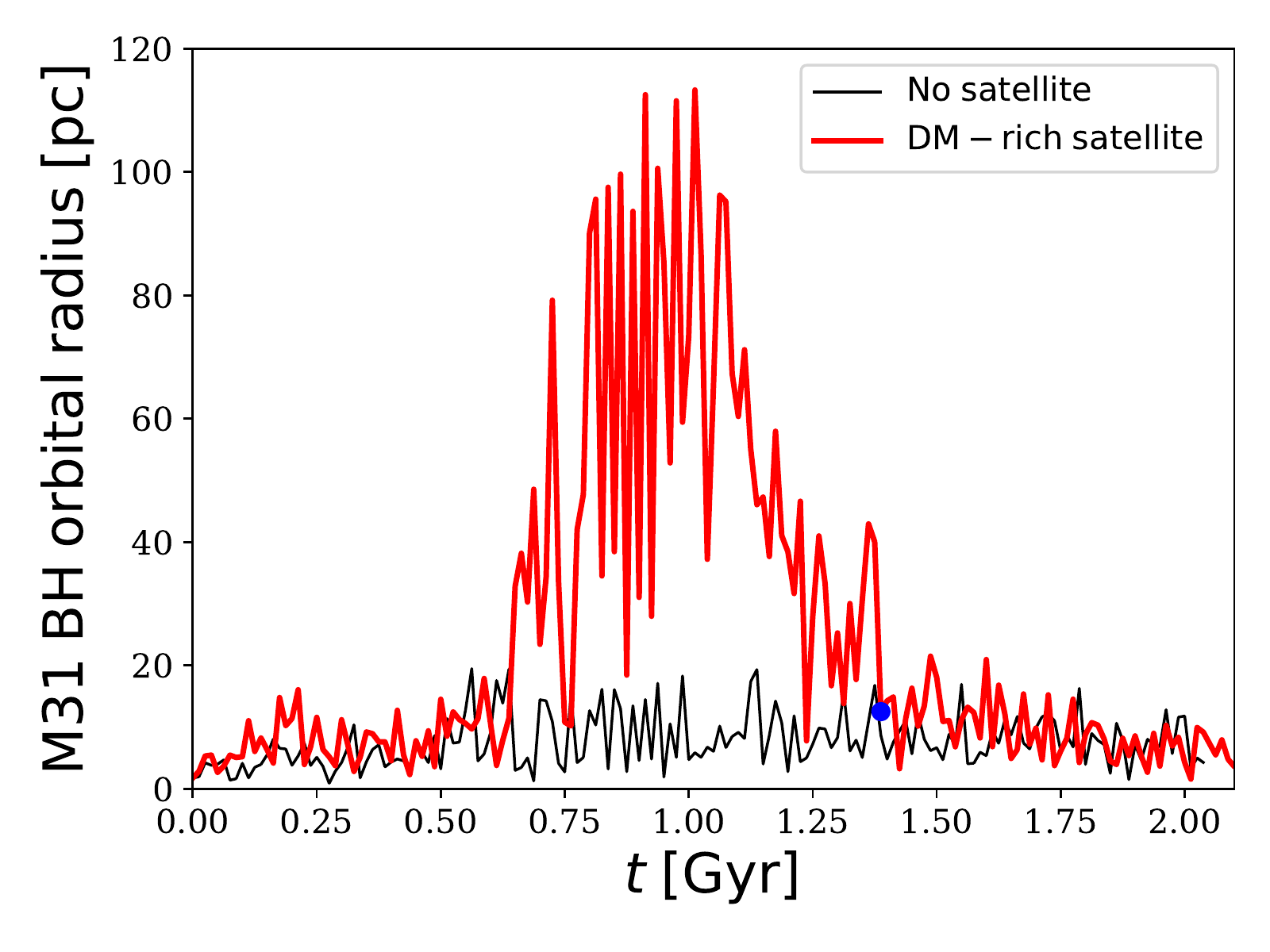}
\caption{{\it Off-centered MBH:} M31 BH orbital radius over 2.1 Gyr. This radius corresponds to the distance between the BH and the mass centre of the M31 stellar component. The MBH of mass $1.5\times10^{8}$ M$_{\sun}$ is initially at the centre of M31. The first passage of the satellite heat the central region and more particularly affect the MBH via dynamical friction. Indeed, the dark matter rich (DM-rich) satellite adds energy to the MBH, causing it to leave the galaxy centre. In the absence of satellite, the MBH remains at the centre of the dwarf galaxy. This scenario ensures the stability of the BH against numerical effects (black curve). However, the first pericentric passage of the satellite in M31 results in a kick of the MBH to hundreds of parsecs from the galaxy centre (red curve)}. 
\label{fgR3}
\end{figure}

We consider the accretion of a dark matter rich satellite by M31, which hosts a central MBH (see details in Table~\ref{tab1}). Dynamical friction induced by the dark matter (DM) field of M31 is responsible for the infall of the satellite. As a result, the central region of the galaxy experiences multiple satellite crossings. The latter heat the central region and more particularly the MBH via dynamical friction. After the first pericentric passage, the dark matter rich satellite adds energy to the BH, causing it to leave the galaxy centre. Fig.~\ref{fgR3} illustrates the orbital radius of a $1.5\times10^{8}$ M$_{\sun}$ MBH, initially at the galaxy centre, over 2.2 Gyr. This radius corresponds to the distance between the BH and the mass centre of the M31 stellar component. In the absence of satellite, the MBH should remain at the centre of the dwarf galaxy. With a particle resolution of $4.4\times10^{4}$ M$_{\sun}$ and a softening length of 2 pc, we cannot resolve properly the BH dynamics below 2 pc. Numerical artifacts amplify the expected Brownian motion of the MBH at the M31 centre (black curve) \citep{2007AJ....133..553M}. However, the first pericentric passage of the satellite in M31 results in a kick of the MBH to hundreds of parsecs from the galaxy centre (red curve), where the BH dynamic is resolved in our simulation. 

\begin{figure}
\centering
\includegraphics[width=0.47\textwidth]{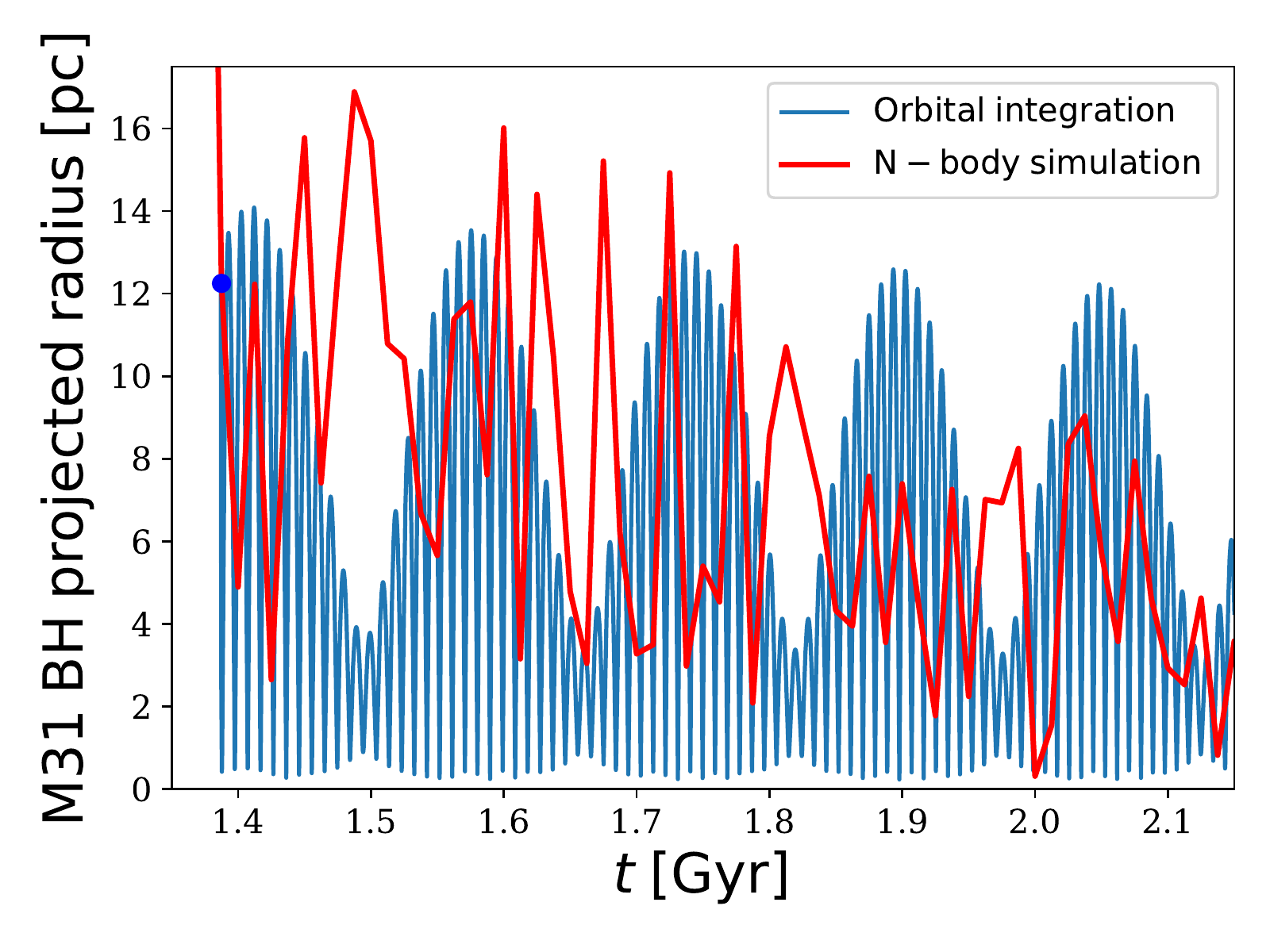}
\caption{{\it Zoom on the BH offset:} M31 BH projected radius from simulation and orbit integration between 1.35 and 2.15 Gyr. We take initial conditions for the MBH at resolved scales (above 10 pc) from the simulation in order to reduce our numerical noise. For our calculation, we considered all M31 components detailed in Table.~\ref{tab1}. The MBH is still offset and is orbiting with a mean projected pericentre of 0.45 pc, which is similar to the value of 0.26 pc derived from observations \citep{1999ApJ...522..772K}. As shown before, the best match between the observed and simulated stream is obtained at 2.1 Gyr, which corresponds to a BH projected pericentre of 0.39 pc for orbital integrations from $t_0=1.38$ Gyr.}
\label{fgR4}
\end{figure}

After having reached its maximum offset, the MBH sinks towards the M31 centre due to dynamical friction. Near the M31 centre, Fig.~\ref{fgR3} depicts a stalling behaviour of the MBH due to numerical effects. As M31 BH is currently offset by 0.26 pc, we want to determine the BH fate in the centre of galaxy (below 2 pc). That is the reason why we integrate the orbits of the MBH forward in time using the galpy package \citep{2015ApJS..216...29B} by taking into account dynamical friction. Via this semi-analytical approach, we avoid numerical effects due to a lack of particle resolution and a softening length. We employ initial conditions  at $t_0=1.38$ Gyr for the MBH at resolved scales (above 10 pc) from the simulation with a satellite in order to reduce our numerical noise. For our calculation, we considered all M31 components detailed in Table.~\ref{tab1}. Fig.~\ref{fgR4} compares the M31 BH projected radius from simulation and orbit integrations between 1.35 and 2.15 Gyr. We have tested different initial times $t_0$ for the orbital integrations of the MBH. According to Figure~\ref{fgR4}, the MBH is still offset and is orbiting with a mean projected pericentre of 0.45 pc, which is similar to the value of 0.26 pc derived from observations \citep{1999ApJ...522..772K}. As shown before, the best match between the observed and simulated stream is obtained at 2.1 Gyr, which corresponds to a BH projected pericentre of 0.39 pc for orbital integrations from $t_0=1.38$ Gyr. The discrepancy between our estimation and observation could be explained by the unavoidable numerical artifacts highlighted previously. Moreover, our result suggests that the MBH is still orbiting at M31 centre and is currently observed at its pericentre if the accreting satellite is responsible for the BH offset. Despite a recent merger with a dark matter rich satellite, we demonstrate that the MBH had sufficient time to come back to M31 centre. Thus, we establish that the infall of the accreting satellite in M31 naturally explains a BH offset by sub-parsecs. At the same time, we also ruled out the Brownian motion of the MBH as the origin of this offset.

\section{Conclusion}
Using high-resolution numerical simulations, we reaffirmed that the accretion of a dark-matter-rich satellite reproduces successfully the tidal features such as the giant stellar stream and the two shells in M31, which hosts a central black hole \citep{2014MNRAS.442..160S}. In this work, we have shown that the heating of the central region of M31 by this dark matter rich satellite via dynamical friction entails a significant MBH offset after the first pericentric passage by using a fully GPU state-of-the-art $N$-body simulation. Using orbital integrations, we highlighted the sinking of the BH towards the parsec scale in M31. The heating by the satellite and the subsequent kick to the central MBH naturally explains a present BH offset by sub-parsecs in M31, detected by \cite{1999ApJ...522..772K}. Our result reinforces the prediction of \cite{2020arXiv200302611B} concerning the furthest distance reached by MBHs in high mass galaxies. Indeed, they pointed out that MBHs are going to have less inertia due to the lower potential in these galaxies. Dynamical perturbations induced by satellite crossings, causing the MBH to vacate the galaxy centre of M31, trigger also a cusp-core transition in the DM halo of M31, generating a core in the presence of a MBH.

\section{Acknowledgments}
We thank the reviewer for their constructive feedback which helped  to  improve  the  quality  of  the  manuscript. We thank also Roya Mohayaee and David Valls-Gabaud for illuminating discussions about M31. We thank Miki Yohei for kindly providing us with the non-public $N$-body code, \textsc{gothic}. We would like also to thank Apolline Guillot, Dante von Einzbern and George Boole for their constructive suggestions on improving the manuscript.


\bsp	
\label{lastpage}

\begin{thebibliography}{2}

\bibitem[\protect\citeauthoryear{Bahcall \& Wolf}{1976}]{1976ApJ...209..214B} Bahcall J.~N., Wolf R.~A., 1976, ApJ, 209, 214

\bibitem[\protect\citeauthoryear{Barth, et al.}{2009}]{2009ApJ...690.1031B} Barth A.~J., Strigari L.~E., Bentz M.~C., Greene J.~E., Ho L.~C., 2009, ApJ, 690, 1031

\bibitem[\protect\citeauthoryear{Boldrini, et al.}{2020a}]{2020arXiv200212192B} Boldrini P., Mohayaee R., Silk J., 2020, arXiv, arXiv:2002.12192

\bibitem[\protect\citeauthoryear{Boldrini, et al.}{2020b}]{2020arXiv200302611B} Boldrini P., Mohayaee R., Silk J., 2020, arXiv, arXiv:2003.02611

\bibitem[\protect\citeauthoryear{Bovy}{2015}]{2015ApJS..216...29B} Bovy J., 2015, ApJS, 216, 29

\bibitem[\protect\citeauthoryear{Comerford \& Greene}{2014}]{2014ApJ...789..112C} Comerford J.~M., Greene J.~E., 2014, ApJ, 789, 112

\bibitem[\protect\citeauthoryear{Fardal, et al.}{2006}]{2006MNRAS.366.1012F} Fardal M.~A., Babul A., Geehan J.~J., Guhathakurta P., 2006, MNRAS, 366, 1012

\bibitem[\protect\citeauthoryear{Fardal, et al.}{2007}]{2007MNRAS.380...15F} Fardal M.~A., Guhathakurta P., Babul A., McConnachie A.~W., 2007, MNRAS, 380, 15

\bibitem[\protect\citeauthoryear{Fardal, et al.}{2013}]{2013MNRAS.434.2779F} Fardal M.~A., et al., 2013, MNRAS, 434, 2779

\bibitem[\protect\citeauthoryear{Ferguson, et al.}{2002}]{2002AJ....124.1452F} Ferguson A.~M.~N., Irwin M.~J., Ibata R.~A., Lewis G.~F., Tanvir N.~R., 2002, AJ, 124, 1452

\bibitem[\protect\citeauthoryear{Font, et al.}{2006}]{2006AJ....131.1436F} Font A.~S., Johnston K.~V., Guhathakurta P., Majewski S.~R., Rich R.~M., 2006, AJ, 131, 1436

\bibitem[\protect\citeauthoryear{Gondolo \& Silk}{1999}]{1999PhRvL..83.1719G} Gondolo P., Silk J., 1999, PhRvL, 83, 1719

\bibitem[\protect\citeauthoryear{Guhathakurta, et al.}{2006}]{2006AJ....131.2497G} Guhathakurta P., et al., 2006, AJ, 131, 2497

\bibitem[\protect\citeauthoryear{G{\"u}ltekin, et al.}{2009}]{2009ApJ...698..198G} G{\"u}ltekin K., et al., 2009, ApJ, 698, 198

\bibitem[\protect\citeauthoryear{Hammer, et al.}{2010}]{2010ApJ...725..542H} Hammer F., Yang Y.~B., Wang J.~L., Puech M., Flores H., Fouquet S., 2010, ApJ, 725, 542

\bibitem[\protect\citeauthoryear{Ibata, et al.}{2001}]{2001Natur.412...49I} Ibata R., Irwin M., Lewis G., Ferguson A.~M.~N., Tanvir N., 2001, Natur, 412, 49

\bibitem[\protect\citeauthoryear{Ibata, et al.}{2004}]{2004MNRAS.351..117I} Ibata R., Chapman S., Ferguson A.~M.~N., Irwin M., Lewis G., McConnachie A., 2004, MNRAS, 351, 117

\bibitem[\protect\citeauthoryear{Ibata, et al.}{2005}]{2005ApJ...634..287I} Ibata R., Chapman S., Ferguson A.~M.~N., Lewis G., Irwin M., Tanvir N., 2005, ApJ, 634, 287

\bibitem[\protect\citeauthoryear{Kirihara, Miki \& Mori}{2014}]{2014PASJ...66L..10K} Kirihara T., Miki Y., Mori M., 2014, PASJ, 66, L10

\bibitem[\protect\citeauthoryear{Kirihara, et al.}{2017}]{2017MNRAS.464.3509K} Kirihara T., Miki Y., Mori M., Kawaguchi T., Rich R.~M., 2017, MNRAS, 464, 3509

\bibitem[\protect\citeauthoryear{Koch, et al.}{2008}]{2008ApJ...689..958K} Koch A., et al., 2008, ApJ, 689, 958

\bibitem[\protect\citeauthoryear{Komossa}{2012}]{2012AdAst2012E..14K} Komossa S., 2012, AdAst, 2012, 364973

\bibitem[\protect\citeauthoryear{Kormendy \& Richstone}{1995}]{1995ARA&A..33..581K} Kormendy J., Richstone D., 1995, 33, 581

\bibitem[\protect\citeauthoryear{Kormendy \& Bender}{1999}]{1999ApJ...522..772K} Kormendy J., Bender R., 1999, ApJ, 522, 772

\bibitem[\protect\citeauthoryear{Kormendy \& Ho}{2013}]{2013ARA&A..51..511K} Kormendy J., Ho L.~C., 2013, 51, 511

\bibitem[\protect\citeauthoryear{Lauer, et al.}{1993}]{1993AJ....106.1436L} Lauer T.~R., et al., 1993, AJ, 106, 1436

\bibitem[\protect\citeauthoryear{Loeb}{2007}]{2007PhRvL..99d1103L} Loeb A., 2007, PhRvL, 99, 041103

\bibitem[\protect\citeauthoryear{McConnachie, et al.}{2003}]{2003MNRAS.343.1335M} McConnachie A.~W., Irwin M.~J., Ibata R.~A., Ferguson A.~M.~N., Lewis G.~F., Tanvir N., 2003, MNRAS, 343, 1335

\bibitem[\protect\citeauthoryear{Merritt \& Milosavljevi{\'c}}{2005}]{2005LRR.....8....8M} Merritt D., Milosavljevi{\'c} M., 2005, LRR, 8, 8

\bibitem[\protect\citeauthoryear{Merritt, Berczik \& Laun}{2007}]{2007AJ....133..553M} Merritt D., Berczik P., Laun F., 2007, AJ, 133, 553

\bibitem[\protect\citeauthoryear{Miki, et al.}{2014}]{2014ApJ...783...87M} Miki Y., Mori M., Kawaguchi T., Saito Y., 2014, ApJ, 783, 87

\bibitem[\protect\citeauthoryear{Miki, Mori \& Rich}{2016}]{2016ApJ...827...82M} Miki Y., Mori M., Rich R.~M., 2016, ApJ, 827, 82

\bibitem[Miki and Umemura(2017)]{2017NewA...52...65M} Miki, Y., Umemura, M.\ 2017.\ GOTHIC: Gravitational oct-tree code accelerated by hierarchical time step controlling.\ New Astronomy 52, 65.

\bibitem[Miki and Umemura(2017)]{2017arXiv171208760M} Miki, Y., Umemura, M.\ 2017.\ MAGI: many-component galaxy initialiser.\ arXiv e-prints arXiv:1712.08760.

\bibitem[\protect\citeauthoryear{Menezes, Steiner \& Ricci}{2014}]{2014ApJ...796L..13M} Menezes R.~B., Steiner J.~E., Ricci T.~V., 2014, ApJL, 796, L13

\bibitem[\protect\citeauthoryear{Menezes, Steiner \& da Silva}{2016}]{2016ApJ...817..150M} Menezes R.~B., Steiner J.~E., da Silva P., 2016, ApJ, 817, 150

\bibitem[\protect\citeauthoryear{Mori \& Rich}{2008}]{2008ApJ...674L..77M} Mori M., Rich R.~M., 2008, ApJL, 674, L77

\bibitem[\protect\citeauthoryear{Power, et al.}{2003}]{2003MNRAS.338...14P} Power C., et al., 2003, MNRAS, 338, 14

\bibitem[\protect\citeauthoryear{Reines, et al.}{2019}]{2019arXiv190904670R} Reines A., Condon J., Darling J., Greene J., 2019, arXiv, arXiv:1909.04670

\bibitem[\protect\citeauthoryear{Sadoun, et al.}{2014}]{2014MNRAS.442..160S} Sadoun R., Mohayaee R., Colin J., 2014, MNRAS, 442, 160

\bibitem[\protect\citeauthoryear{Shen, et al.}{2019}]{2019ApJ...885L...4S} Shen Y., Hwang H.-C., Zakamska N., Liu X., 2019, ApJL, 885, L4

\bibitem[\protect\citeauthoryear{Sundararajan, Khanna \& Hughes}{2010}]{2010PhRvD..81j4009S} Sundararajan P.~A., Khanna G., Hughes S.~A., 2010, PhRvD, 81, 104009

\bibitem[\protect\citeauthoryear{Tremmel, et al.}{2018}]{2018ApJ...857L..22T} Tremmel M., Governato F., Volonteri M., Pontzen A., Quinn T.~R., 2018, ApJL, 857, L22

\bibitem[\protect\citeauthoryear{Volonteri \& Perna}{2005}]{2005MNRAS.358..913V} Volonteri M., Perna R., 2005, MNRAS, 358, 913

\bibitem[\protect\citeauthoryear{White \& Rees}{1978}]{1978MNRAS.183..341W} White S.~D.~M., Rees M.~J., 1978, MNRAS, 183, 341

\bibitem[\protect\citeauthoryear{White \& Frenk}{1991}]{1991ApJ...379...52W} White S.~D.~M., Frenk C.~S., 1991, ApJ, 379, 52

\end{thebibliography}
\end{document}